# A Wide Temperature Range Unified Undoped Bulk Silicon Electron and Hole Mobility Model

Prabjot Dhillon and Hiu Yung Wong, *Senior Member, IEEE*

*Abstract*— For decades, there is no unified model developed for Silicon carriers from 4K to above room temperature. In this paper, a unified undoped Silicon low field and high field mobility model for both electron in the <100> and <111> directions from 8K to 300K and 430K, respectively, and hole in the <100> direction from 6K to 430K is proposed and calibrated to the experiment. It is found that the Canali high field saturation model is sufficient to fit the <111> experimental data but not the <100> data due to the anisotropy-induced plateaus and negative differential velocity. Therefore, a modified Farahmand model is used. To allow parameter interpolation in anisotropic simulation, the modified Farahmand model is also calibrated for the <111> direction. The model is then used to predict the mobility of electrons and holes in undoped Si at 4K, which can be served as the initial calibration parameters when reliable experimental data is available for the TCAD model development.

## I. Introduction

CRYOGENIC microelectronics, in particular Silicon-based Complementary Metal Oxide Semiconductor (Cryo-CMOS), are becoming more important due to their applications in quantum computing interface [1][2], space exploration [3], and high-performance servers [4]. Cryo-CMOS needs to operate at 4.2K or even lower temperatures to reduce noise and latency in the quantum computer interface. The bulk [5], SOI [2], and FinFET [6] devices are all possible candidates. The ultimate goal of qubit-CMOS co-integration on a single chip requires the cryo-CMOS to be optimized so that the heat generated can also be minimized [2]. Therefore, accurate models and parameters are very important to enable the full deployment of cryo-CMOS.

There are many studies on cryo-CMOS modeling at the compact model level [5][7]-[12]. Ref. [7] studied the modeling of device linear region mobility from 4.2K to 300K. Ref.'s [8] and [5] proposed CMOS compact models down to 77K and 4.2K, respectively. Ref. [9] also discussed the low field and high field mobility models in cryoCMOS devices. Refs. [10] and [11] studied the free carrier mobilities. Ref. [11] studied the compact modeling of FinFET while [12] studied the modeling of 0.18µm technology node planar CMOS.

There are also efforts in modeling the cryo-CMOS at the Technology Computer-Aided Design (TCAD) level, including incomplete ionization, abnormal subthreshold slope, and high field saturation models [13]-[16]. However, modeling of electron mobility at cryogenic temperature has not been successful due to its negative differential velocity. Ref. [14] shows that the Canali high field saturation model [17], which is commonly used for room temperature, gives absurd results below 20K and it is improved by using the Selberherr model [2]. Although the high field region can be captured by the Selberherr model, the medium and lower field parts below 77K cannot be modeled well. Firstly, it cannot capture the low field mobility well, and secondly, it cannot model the negative differential velocity. Therefore, a new low field and high field mobility model is required to enable accurate TCAD simulations and also serve as a basis to further develop more accurate compact models for circuit simulations.

Wide temperature range Si electron mobility experimental data are not widely available and one of the most complete sets is reported in [18][19] which contains 8K-300K <111> direction and 8K-430K <100> direction for electron and 6K-430K for <100> hole. We propose to use the modified Farahmand model in [20] to model *all* experimental data in [18] and [19]. The modified Farahmand, which is derived from [21] and [22] in [20], can capture negative differential velocity but has only been demonstrated in III-V materials at room temperatures. We derived smooth functions to model the temperature-dependent parameters in this model so as to predict the mobilities at 4K.

## II. Modeling and Calibration

Usually, in TCAD, the low field mobility is first calculated by combining the temperature-dependent undoped bulk mobility model, $\mu_{bulk}$, with other mobility models using Matthiessen's rule [20],

$$\frac{1}{\mu_{low}} = \frac{1}{\mu_{bulk}} + \frac{1}{\mu_{doped}} + \frac{1}{\mu_{SC}} + \cdots \quad (1)$$

where $\mu_{low}$ is the overall low-field mobility, $\mu_{doped}$ is the impurity dependent mobility (e.g. [23]), and $\mu_{SC}$ is the surface

Manuscript received Oct. 7, 2021. (Corresponding author: Hiu Yung Wong). This material is based upon work supported by the National Science Foundation under Grant No. 2046220.

Prabjot Dhillon and Hiu Yung Wong are with the M-PAC Lab, Electrical Engineering Department, San Jose State University, San Jose, CA, USA (E-mail: hiuyung.wong@sjsu.edu).

scattering mobility, which can be due to various scattering mechanisms such as surface roughness and surface phonon scatterings (e.g. [24]) or remote Coulomb scattering in high-κ dielectric (e.g. [25]). *Such a framework allows the flexibility to add novel scattering mechanisms in novel transistors.*

The final field-dependent mobility is then evaluated based on $\mu_{low}$ by using various high field saturation models (e.g. [17][21]). In this paper, the mobility model proposed has a separable $\mu_{low}$ part accounting only for the undoped bulk mobility (i.e. the first term in (1)) and is readily integrated into any TCAD simulator.

As a first step of the calibration, carrier velocity, $v$, vs. electric field, $E$, experimental data are digitized from [18] (for hole <110> and electron <111>) and [19] (for electron <100>) and the effective mobility, $\mu$, is calculated as $\mu = v/E$. To fit the electron <111> data, it is found that it is adequate to use the Canali model [17],

$$\mu = \mu_{low}\left(1 + \left(\frac{\mu_{low}E}{v_{sat}}\right)^\beta\right)^{-\frac{1}{\beta}} \quad (2)$$

where $\mu_{low}$, $v_{sat}$, and $\beta$ are temperature-dependent parameters. We have derived a unified set of temperature-dependent parameters and their equations are shown in Table I. Note that for the Canali model, $\beta_{sat}=1$, $T_0=130$, b $=3$, $\beta_0=0.4$. Fig. 1 shows the calibration results of the <111> electron mobility. It shows that it can capture a wide temperature range. Since $\mu_{low}$ and $v_{sat}$ have direct physical meanings, they are calibrated

TABLE I
TEMPERATURE DEPENDENT PARAMETERS OF FARAHMAND MODEL
(WIDE TEMPERATURE RANGE)

**Equations and Parameters**

| | |
|---|---|
| $\mu = \dfrac{\mu_{low}+\mu_1\left(\frac{E}{E_0}\right)^\alpha + v_{sat}\frac{E^{\beta-1}}{E_1^\beta}}{1+\gamma\left(\frac{E}{E_0}\right)^\alpha + \left(\frac{E}{E_1}\right)^\beta}$ | (3) [20] [21] |
| $\mu_{low} = A\,T^{-\gamma}$ | (4) [18] |
| $A = 2.36e7, 2.36e7, 1.35e8$; $\gamma = 1.7, 1.7, 2.2$ | |
| $v_{sat} = v^*/\left(1 + Ce^{\frac{T}{\Theta}}\right)$ | (5) [18] |
| $v^* = 2.4e7, 2.4e7, 2.3e7$; $C = 0.8, 0.8, 1.1$; $\Theta = 600, 600, 600$ | |
| $\mu_1 = \left(1 - \tanh((T - T_0)/T_1)\right)/(2k)$ | (6) |
| $T_0 = 50, 50, 15$; $T_1 = 65, 65, 100$; $k = 2900, 2900, 3000$ | |
| $\alpha = \alpha_{sat}(T/T_0)/(1 + (T/T_0)^a)^{1/a} + \alpha_0$ | (7) |
| $\alpha_{sat} = 2.5, 2.54, 2.54$; $T_0 = 80, 98, 98$; $a = 6, 8, 8$; $\alpha_0 = 1.05, 1, 1$ | |
| $\beta = \beta_{sat}(T/T_0)/(1 + (T/T_0)^b)^{1/b} + \beta_0^*$ | (8) |
| $\beta_{sat} = 2.8, 2.5, 2.5$; $T_0 = 85, 120, 120$; b $= 5, 4, 4$; $\beta_0 = 1.8, 2.2, 2.2$ | |
| $\gamma = \left(k\,\tanh((T - T_0)/T_1) + 1\right) + \gamma_0$ | (9) |
| $k = 0.6, 0.59, 1.64$; $T_0 = 50, 50, 47.81$; $T_1 = 50, 4, 19.5$; $\gamma_0 = 1.7, 1.7, 1.47$ | |
| $E_0 = E_{sat}(\tanh((T - T_3)/T_0) + 1)(\tanh((T/T_1)^p)) \times (1 - A\exp(-|T - T_2|/s))$ | (10) |
| $E_{sat} = 800, 570, 300$; $T_0 = 33.3, 60.3, 83$; $T_1 = 5, 18, 120$; $T_2 = 125, 123, 160$; $T_3 = 125, 123, 90$; $p = 2, 1, 0.8$; $A = 2, 1, -0.4$; $s = 1.67, 6, 33.3$ | |
| $E_1 = e_1 e^{\frac{e_2 T}{T_1}}(\tanh((T - T_0)/T_1) - 1) + (I_1 T + I_2)(\tanh((T - T_0)/T_1) + 1)$ | (11) |
| $e_1 = -1.725, -35, -28$; $e_2 = 0.0441, 0.0158, 0.0157$; $I_1 = 1.5, 0.5, 0.7$; $I_2 = 340, 475, 240$; $T_0 = 130, 200, 190$; $T_1 = 1, 1, 1$ | |

Parameters are given in the order of <111> electron, <100> electron and <100> hole. $T$ is the temperature in kelvin. $T_0$, $T_1$, $T_2$, and $T_3$ have the unit of kelvin. Centimeter is used as the length unit and the rest are in SI. The units of the rest of the parameters have the natural units to match the dimensions. * $\beta$ for the Canali model uses the same equation but the parameters are shown in the main text after (2). Eq. (6) to (11) are developed in this paper.

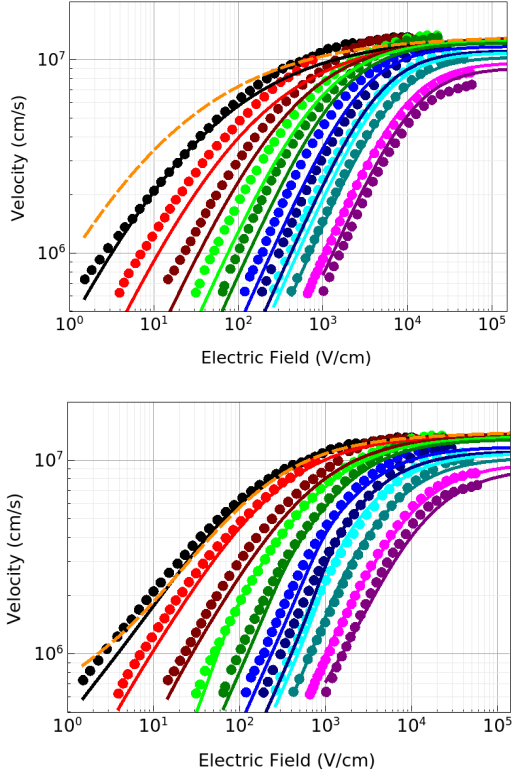

Fig. 1. <111> electron velocity as a function of electric field at 8K, 20K, 45K, 77K, 110K, 160K, 220K, 245K, 300K, 370K, and 430K. Continuous lines: Model; Dots: Experiment [18]. Orange dashed line: 4K predicted by the models. Top: Canali model [17]. Bottom: Modified Farahmand.

using the well-known equations in [18] with parameters such that the temperature-dependent low field mobility and saturation velocity matches the experimental results. Electron $\mu_{low}$ and $v_{sat}$ are expected to be isotropic in silicon and thus they are the same in both <100> and <111> directions. However, at low temperature and medium electric field, the negative differential velocity with respect to electric field appears in <100> direction because electrons transporting at the transverse valleys are heated up and repopulate in the heavier longitudinal value [18]. This does not happen in <111> due to symmetry and does not happen at the low field due to lack of intervalley scattering and neither at the high field due to sufficient intervalley scattering. This cannot be modeled by the Canali model. Therefore, we used a modified Farahmand model from [20]

$$\mu = \dfrac{\mu_{low}+\mu_1\left(\frac{E}{E_0}\right)^\alpha + v_{sat}\frac{E^{\beta-1}}{E_1^\beta}}{1+\gamma\left(\frac{E}{E_0}\right)^\alpha + \left(\frac{E}{E_1}\right)^\beta} \quad (3)$$

where $\mu_1$, $\alpha$, $\beta$, $\gamma$, $E_0$, and $E_1$ are parameters and we have derived the parameter equations and calibrated the parameters as shown in Table I. $\mu_{low}$ and $v_{sat}$ share the same values as in the Canali model. As shown in Fig. 2, the model can capture the transition from a deep slope to a plateau at a medium electric field in the curves when the temperature is reduced to 45K.

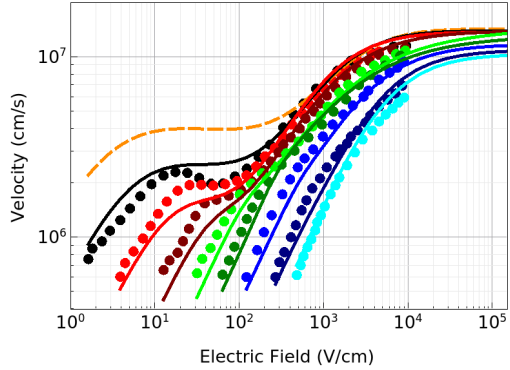

Fig. 2. <100> electron velocity as a function of electric field at 8K, 20K, 45K, 77K, 110K, 160K, 245K, and 300K. Continuous lines: Model; Dots: Experiment [19]. Orange dashed line: 4K predicted by the Modified Farahmand model.

To allow interpolation between <100> and <111> directions in TCAD simulation, the <111> electron mobility is also calibrated using the Farahmand model and shown in Fig. 1. Overall, it gives a better fitting compared to using the Canali model, particularly in the high field part.

The hole velocity also exhibits a flat plateau at low temperature (<30K) and medium electric field in the <100> direction. By using the same set of equations, the experimental data can be fitted well also (Fig. 3). Note that there is no hole velocity saturation data in the literature below 300K. The $v_{sat}$ parameters for the hole in (5) in Table I are derived so that it matches high-temperature experimental data [17] while giving a smooth curve. Therefore, hole parameters in (4) are obtained as a part of the overall calibration. Also note that at T<24K, the hole low field mobilities start deviating from theory. This is due to the background impurities (<$10^{12}$cm$^{-3}$) [18]. Hence, to fit the 6K data and extrapolate the 4K curve, (3) is not used but $5\times10^5$cm$^2$/Vs and $7\times10^5$cm$^2$/Vs extracted from experimental data with impurities <$10^{12}$cm$^{-3}$ are used.

### III. DISCUSSION

In the model development and calibration process, strategies

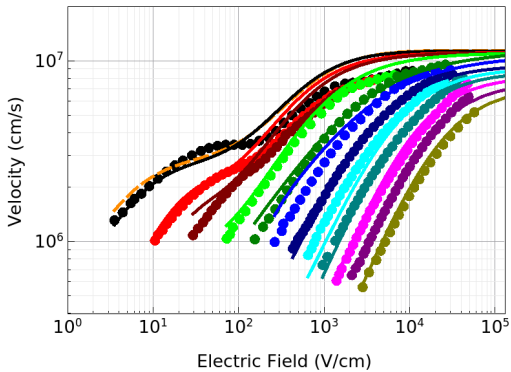

Fig. 3. <100> hole velocity as a function of electric field at 6K, 24K, 30K, 45K, 77K, 110K, 160K, 200K, 245K, 300K, 370K, and 430K. Continuous lines: Model; Dots: Experiment [18]. Orange dashed line: 4K predicted by the Modified Farahmand model.

TABLE II
4K FARAHMAND MODEL PARAMETERS FOR ELECTRONS AND HOLES

| Type | $\mu_{low}$# | $v_{sat}$* | $\mu_1$ | $\alpha$ | $\beta$ | $\gamma$ | $E_0$ | $E_1$ |
|---|---|---|---|---|---|---|---|---|
| e111 | 2.24 | 1.33 | 1408 | 1.17 | 1.93 | 2.06 | 0.72 | 4.12 |
| e100 | 2.24 | 1.33 | 2333 | 1.10 | 2.28 | 2.11 | 4.72 | 76.6 |
| h100 | 0.7 | 1.09 | 1664 | 1.10 | 2.28 | 0.87 | 4 | 60 |

#×$10^6$. *×$10^7$. All units are the same as in Table I.

are used to avoid overfitting and to capture the physics as much as possible. Firstly, the Farahmand model which has been proven to be useful in modeling negative differential velocity in III-V materials is used. This model is also proven to be suitable for TCAD simulation. Secondly, $\mu_{low}$ and $v_{sat}$ which have direct physical meanings are kept as they are and calibrated to experimental data whenever available. Thirdly, differentiable equations are used to model the temperature-dependent parameters, and efforts are made to ensure them to be monotonic. The ability of the Farahmand model to model the <111> electron experimental result as well as the Canali model in Fig. 1 shows that it probably is not overfitted. With confidence in the parameter equations, the 4K velocity vs. field curves are predicted and plotted as the orange dashed curves in Figs. 1-3 and the parameters are shown in Table II. Although there is a difference between the 4K curves predicted by the Farahmand model and the Canali model in Fig. 1, the difference is less than 20% for $E >$100V/cm.

In TCAD, the proposed model will be very useful for self-heating simulations where the device temperature is non-uniform spatially. For a device operating under normal conditions (e.g. interface electronics for quantum computing), it is expected that the temperature range within one single device should not exceed the 4K-40K range. Therefore, another set of parameters with the same models is proposed to provide a better fit to the experimental data ≤ 45K. Table III shows the parameters. Fig. 4 and Fig. 5 show the fitting results for <100> electron and <100> hole mobilities, respectively. It is found that the fitting can much better if it is not required to fit the whole range of temperature to up to 400K. Particularly, the negative differential velocity of <100> electron at 8K and the flat plateau of <100> hole at 6K can be captured very well.

If the device operates in a very wide temperature range (e.g. 4K to 400K) within the device dimension, although the parameters in Table I do not capture the negative differential velocity as well, this part of mobility is expected not to be important as the device must be in the high field regime to have serious self-heating.

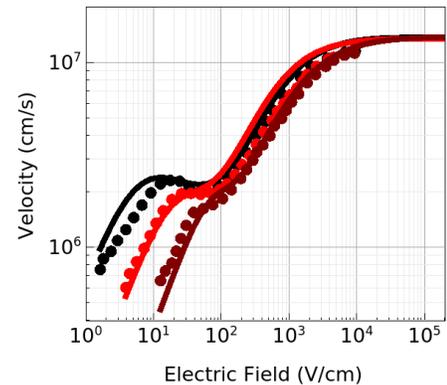

Fig. 4. <100> electron velocity as a function of electric field at 8K, 20K, and 45K using the parameters in Table III. Continuous lines: Model; Dots: Experiment [19].

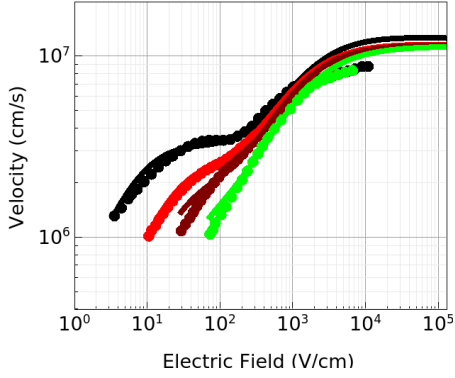

Fig. 5. <100> hole velocity as a function of electric field at 6K, 24K, 30K, and 45K using the parameters in Table III. Continuous lines: Model; Dots: Experiment [18].

Finally, we will discuss the parameter fitting process.

Firstly, at low electric fields, $\mu_{low}$ dominates in (3) as $E$ is small. At medium electric fields, the second term in the numerator of (3) needs to dominate so to capture the negative or near-zero differential velocity. This depends on $E_0$, $\mu_1$, and $\gamma$, and also on the ratio between $\mu_1$, and $\gamma$. At high electric fields, the third term needs to dominate so that $\mu = \frac{v_{sat}}{E}$ at a very high electric field by definition. This is dependent on $E_1$. In order to make the third term dominate over the second term, $\beta$ needs to be larger than $\alpha$. This will also determine the flatness of the velocity curves at high electric fields. Each temperature is fitted manually by adjusting the parameters in (3) based on this methodology. In this process, care is taken to make sure there is a smoothness in the parameter values across various temperatures.

After the individual temperature is fitted well with all parameters in (3) showing a smooth trend across various temperatures, more fundamental parameters in (7) to (11) are fitted so that they produce the parameter values fitted for (3).

The fitting is gauged by 3 means. Firstly, it can capture the shape (e.g. negative differential velocity) and trend of the experimental curves well. Secondly, the fundamental parameters in (7) to (11) have reasonable values among electrons and holes (most of them are in the same order of magnitude). Thirdly, the average fitting error, when compared to the experiment, is less than 10%. For example, the average error of Fig. 4 and Fig. 5 are 3.8% and 1.4%, respectively.

## IV. CONCLUSIONS

We have developed and calibrated a unified field-dependent and temperature-dependent mobility model in Silicon for the electron in the <100> and <111> directions and the hole in the <100> direction. The relationship between the carrier mobilities and the electric field is based on the Farahmand model and shown in the carrier velocity-vs-electric field plots. The model is developed such that it can capture the negative differential velocity and the functions are smooth to enable the prediction of the carriers at 4K. This will allow self-heating simulation in TCAD in the cryogenic regime.

TABLE III
TEMPERATURE DEPENDENT PARAMETERS OF FARAHMAND MODEL
(NARROWER TEMPERATURE RANGE)

**Equations and Parameters**

| | |
|---|---|
| $\mu = \dfrac{\mu_{low} + \mu_1 \left(\frac{E}{E_0}\right)^\alpha + v_{sat}\frac{E^{\beta-1}}{E_1^\beta}}{1 + \gamma\left(\frac{E}{E_0}\right)^\alpha + \left(\frac{E}{E_1}\right)^\beta}$ | (3) [20] [21] |
| $\mu_{low} = A\, T^{-\gamma}$ | (4) [18] |
| $A = 2.36e7, 2.36e7, 1.35e8$; $\gamma = 1.7, 1.7, 2.2$ | |
| $v_{sat} = v^*/\left(1 + C e^{\frac{T}{\Theta}}\right)$ | (5) [18] |
| $v^* = 2.4e7, 2.4e7, 2.3e7$; $C = 0.8, 0.8, 1.1$; $\Theta = 600, 600, 600$ | |
| $\mu_1 = \left(1 - \tanh((T - T_0)/T_1)\right)/(2k)$ | (6) |
| $T_0 = 50, 50, 15$; $T_1 = 65, 65, 100$; $k = 2900, 2900, 3000$ | |
| $\alpha = \alpha_{sat}(T/T_0)/(1 + (T/T_0)^a)^{1/a} + \alpha_0$ | (7) |
| $\alpha_{sat} = 2.5, \mathbf{3}, \mathbf{1.4}$; $T_0 = 80, \mathbf{110}, \mathbf{170}$; $a = 6, \mathbf{10}, \mathbf{15}$; $\alpha_0 = 1.05, \mathbf{1.21}, \mathbf{1.14}$ | |
| $\beta = \beta_{sat}(T/T_0)/(1 + (T/T_0)^b)^{1/b} + \beta_0$ | (8) |
| $\beta_{sat} = 2.8, \mathbf{5.5}, \mathbf{1.1}$; $T_0 = 85, \mathbf{230}, \mathbf{140}$; $b = 5, \mathbf{2}, \mathbf{1.1}$; $\beta_0 = 1.8, 2.36, 2.25$ | |
| $\gamma = \left(k\, \tanh((T - T_0)/T_1) + 1\right) + \gamma_0$ | (9) |
| $k = 0.6, 0.59, \mathbf{2}$; $T_0 = 50, 50, \mathbf{80}$; $T_1 = 50, 4, \mathbf{58.8}$; $\gamma_0 = 1.7, \mathbf{1.84}, \mathbf{1.5}$ | |
| $E_0 = E_{sat}\left(\tanh((T - T_3)/T_0) + 1\right)\left(\tanh((T/T_1)^p)\right) \times (1 - A\exp(-|T - T_2|/s))$ | (10) |
| $E_{sat} = 800, \mathbf{560}, \mathbf{535.6}$; $T_0 = 33.3, \mathbf{64}, \mathbf{57.7}$; $T_1 = 5, \mathbf{32}, \mathbf{73.2}$; $T_2 = 125, \mathbf{123}, 160$; $T_3 = 125, \mathbf{110}, \mathbf{105.3}$; $p = 2, 1, \mathbf{0.585}$; $A = 2, 1, 0$; $s = 1.67, \mathbf{18}, \mathbf{100}$ | |
| $E_1 = e_1 e^{\frac{e_2 T}{T_1}}\left(\tanh((T - T_0)/T_1) - 1\right) + (I_1 T + I_2)(\tanh((T - T_0)/T_1) + 1)$ | (11) |
| $e_1 = -1.725, \mathbf{-20}, \mathbf{-46.5}$; $e_2 = 0.0441, \mathbf{0.033}, \mathbf{0.017}$; $I_1 = 1.5, 0.5, 0.7$; $I_2 = 340, 475, 240$; $T_0 = 130, \mathbf{155}, 190$; $T_1 = 1, 1, 1$ | |

This table uses the same explanations as in Table I. The equations are the same as those in Table III. Some parameters for <100> electron and <100> hole are modified for better fittings for T≤45K and highlighted in bold. The parameters here should only be used for T≤45K.